\documentclass[11pt]{article}
\usepackage{hyperref}
\pdfoutput=1
\begin{document}
\title{Direct Numerical Simulation of 3D Salt Fingers:\\ From Secondary Instability to Chaotic Convection}
\author{\vspace{11pt} Julian A. Simeonov$^{1}$, Melvin E. Stern$^{2}$ and Timour Radko$^{3}$ \\
\\\vspace{6pt} ${^1}$Naval Research Laboratory, Stennis Space Center 
\\\vspace{6pt} ${^2}$Department of Oceanography, Florida State University
\\\vspace{6pt} ${^3}$Department of Oceanography, Naval Postgraduate School}
\date{} 
\maketitle
\begin{abstract}
The amplification and equilibration of three-dimensional salt fingers in unbounded uniform vertical gradients of temperature and salinity is modeled with a Direct Numerical Simulation in a triply periodic computational domain. A fluid dynamics video of the simulation shows that the secondary instability of the fastest growing square-planform finger mode is a combination of the well-known vertical shear instability of two-dimensional fingers \cite{Holyer1984} and a new horizontal shear mode.
\end{abstract}

\section{Introduction}
Salt fingers are a centimeter scale convection process which takes place in layers that are warm and salty at the top and cold and fresh at the bottom. Such configurations arise commonly in a variety of larger scale oceanic processes such as currents, eddies and frontal instabilities that bring together waters with different temperature and salinity. The salt fingers therefore play an important role in the mixing of oceanic water masses. The primary salt finger mode is an array of depth independent convection cells \cite{Stern1960} consisting of rising cold and fresh water and sinking warm and salty water. The instability is driven by buoyancy gain due to differential diffusion of heat and salt across the convection cell boundaries and is limited by viscous momentum dissipation. The buoyancy gain is further offset by the vertical displacement of the mean (stable) density gradient. Thus, the preferred horizontal wavelength H of the instability is determined by the interplay of these three effects - very thin fingers are suppressed by strong viscous damping, while very wide ones do not gain enough buoyancy through diffusion to overcome the stabilizing effect of the mean density gradient.\\	At Prandtl number Pr=7, two-dimensional salt fingers are unstable to a super-exponentially growing vertical shear mode \cite{Stern2005} which extracts energy by tilting the fingers. The purpose of the present \href{http://hdl.handle.net/1813/13966}{video} is to provide further details on the secondary instability of 3D fingers. The video is generated by spectral Direct Numerical solutions of the non-dimensional Navier-Stokes, and heat and salt conservation equations (see \cite{Stern2001} for further details). The simulation parameters are density ratio R=3, diffusivity ratio $\tau$ = 1/24 and Prandtl number Pr = 7 for which the preferred non-dimensional horizontal wavelength is H=7.48; the domain size is 4-by-4-by-10 H. The initial condition consisted of the fastest growing salt fingers (wavelength H) of square horizontal planform and some random noise to break the initial symmetry. Noting that salinity is an excellent tracer due to its low diffusivity, we visualize the salt finger dynamics with two salinity isosurfaces +10 (yellow) and -10 (light blue). The video begins with the initial exponential growth of the primary mode as evident in the initial time variation of the Nusselt number Nu. At time t = 170 min, the development of a 3D instability interrupts the growth of the primary fingers. The video shows that the secondary instability is a horizontal flow with both vertical and horizontal shear variation. Unlike the two-dimensional case, this shear combination produces extensive straining of the primary mode so that the latter breaks down into blobs with a typical width several times smaller than H. In the final stage of the equilibration 170 min $<$ t $<$ 250 min, the small blobs merge into larger plumes until the convection reaches a statistical equilibrium. Similar small-domain DNS \cite{Stern2001} determined the statistically averaged salt finger heat and salt fluxes for a range of density ratios. These flux laws have been very useful in modeling the amplification of longer vertical wavelengths \cite{Stern2001,Radko2003,Simeonov2007} by the salt finger buoyancy fluxes. Such large scale instabilities tend to produce well-mixed convecting layers in an initially homogeneous field of salt fingers.


\begin{thebibliography}{9}

\bibitem{Holyer1984}
Holyer, J. Y., 1984 
The stability of long steady, two dimensional salt fingers. 
J. Fluid Mech. 
{\bf147}, 169-185.
\bibitem{Stern1960}
Stern, M. E., 1960
The salt fountain and thermohaline convection. 
Tellus 
{\bf12}, 172-175.
\bibitem{Stern2005}	
Stern, M. E. {\&} J. Simeonov, 2005 
The secondary instability of salt fingers. 
J. Fluid Mech. 
{\bf533}, 361-380.
\bibitem{Stern2001}	
Stern, M. E., T. Radko {\&} J. Simeonov, 2001 
Salt fingers in an unbounded thermocline. 
J. Mar. Res. 
{\bf59}, 355-390.
\bibitem{Radko2003}	
Radko, T., 2003 
A mechanism for layer formation in a double-diffusive fluid. 
J. Fluid Mech. 
{\bf497}, 365-380.

\bibitem{Simeonov2007}
Simeonov, J. A. {\&} M. E. Stern, 2007
Equilibration of two-dimensional double-diffusive intrusions
J. Phys. Oceanogr.
{\bf37}, 625-641
\end{thebibliography}
\end{document}